\documentclass{article}
\usepackage{amsthm}
\usepackage{amsmath}
\usepackage{amsfonts}
\usepackage{natbib}
\usepackage[colorlinks,citecolor=blue,urlcolor=blue,filecolor=blue,backref=page]{hyperref}
\usepackage{graphicx}
\usepackage{bm}
\usepackage{multirow}
\usepackage{siunitx}
\usepackage{color}
\usepackage{caption}
\usepackage{subcaption}

\begin{document}
\title{Bayesian Diagnostics for Chain Event Graphs}
\author{Rachel L. Wilkerson and Jim Q. Smith}
\maketitle	

		\begin{abstract}
			Chain event graphs have been established as a practical Bayesian graphical tool. While bespoke diagnostics have been developed for Bayesian Networks, they have not yet been defined for the statistical class of Chain Event Graph models. Mirroring the methodology of prequential diagnostics for Bayesian Networks, in this paper we design a number of new Bayesian diagnostics for this new class. These can be used to check whether a selected model--presumably the best within the class--captures most of the salient features of the observed process. These are designed to check the continued validity of a selected model as data about a population is collected. A previous study of childhood illness in New Zealand illustrates the efficacy of these diagnostics. A second example on radicalisation is used as a more expressive example.
		\end{abstract}

	\section{Introduction }
	
	
	Chain Event Graphs (CEGs) are a useful graphical model representation. They generalise the class of Bayesian Networks (BNs), representing  context-specific independence and graphical asymmetry. Furthermore it can be argued that because they are drawn from a tree-based structure, CEGs allow a more natural way to express a series of unfolding events \citep{Shafer1996}.

	As with other graphical models, CEGs are then populated with distributions, often inferred by data. Typically, these parameters of the distribution can be updated sequentially as more data becomes available. 
	In this setting the routine use of diagnostics is essential.
	They reveal problematic structural elements, expose when changes in the data are no longer compatible with the model, or alternatively demonstrate its plausibility.
	
	
	
	
	Within the Bayesian paradigm prequential diagnostics of \citet{Dawid1984}  have proved particularly useful and simple to apply.
	These examine the one-step ahead forecasts of each subsequent observation in a dataset to determine the compatibility of the model with the data. In particular, prequential diagnostics determine how well the model predicts future data based on past performance \citep{Dawid1992}. These have been used successfully to provide diagnostics for the Bayesian Network class \citep{diagnostics}. 
	
	Prequential diagnostics have since been extended to other graphical models \citep{Costa2015}. Here we extend them to CEGs. The prequential approach is especially attractive for use with this class since its focus is on a model's ability to forecast the future development of a unit in the population given the past. This harmonises beautifully with the type of modelling structure expressed by a CEG which encodes possible future pathways for each unit.
	
	In this paper we describe the suite of diagnostic monitors developed for detecting ill-fitting CEGs. 
	Section~\ref{sec:ceg} explains the meaning and estimation of the Chain Event Graphs and their derivation from the staged trees. In Section~\ref{sec:diag}, we review the prequential diagnostics for the Bayesian Network (BN) and define analogous diagnostics for the CEG in Section~\ref{sec:cegdiag}. Section~\ref{sec:ex} shows the diagnostics applied to two different examples. First, the Christchurch Health and Development Study (CHDS) example shows the process of households circumstances that may result in a child being admitted to the hospital. This example demonstrates the ability of the diagnostic monitors to differentiate between candidate models including a BN and two CEGs. 
	The  radicalisation data shows how individuals in a prison may choose to engage in radical activity. Our second example shows how these diagnostics improve model interpretability  as we begin to scale the CEG. Together, these examples demonstrate how the diagnostics highlight misspecifications in the structure.

	\section{Chain Event Graphs, their meaning and estimation}\label{sec:ceg}
	\subsection{Christchurch data set}
	
	In this paper we consider two examples to illustrate our methodology.
	The first has the advantage that it has been subject to various different CEG models and so is already well studied, see \citet{Barclay2015, Cowell2014, Barclay2015}.
	The study was conducted at the University of Otago, New Zealand \citep{Fergusson1986a}. It encompassed a five year longitudinal study of several explanatory variables including: 
	
	\begin{itemize}
		\item $X_\text{s}$: Family social background, a categorical variable differentiating between high and low levels according to educational, socio-economic, ethnic measures, and information about the children's birth.
		\item $X_\text{e}$: Family economic status, a categorical variable distinguishing between high and low status with regard to standard of living.
		\item $X_\text{l}$: Family life events, a categorical variable signalising the existence of low (0 to 5 events), average (6 to 9 events) or high (10 or more events) number of stressful events faced by a family over the five years.
		\item $X_\text{h}$: Hospital admissions, a binary variable indicating whether or not a child in the household was hospitalised.
	\end{itemize}
	
	The aim of the CHDS study was to better understand how the different variables above might relate to one another. Previous studies of the CHDS data demonstrated the flexibility and expressiveness of the CEG model over the BN \citep{Barclay2013}.
	We will demonstrate below how the diagnostics we develop here pinpoint exactly how the CEG structure can model the processes better than a BN.

	\begin{figure}[h] 
		\centering
		\includegraphics[width=4.5in]{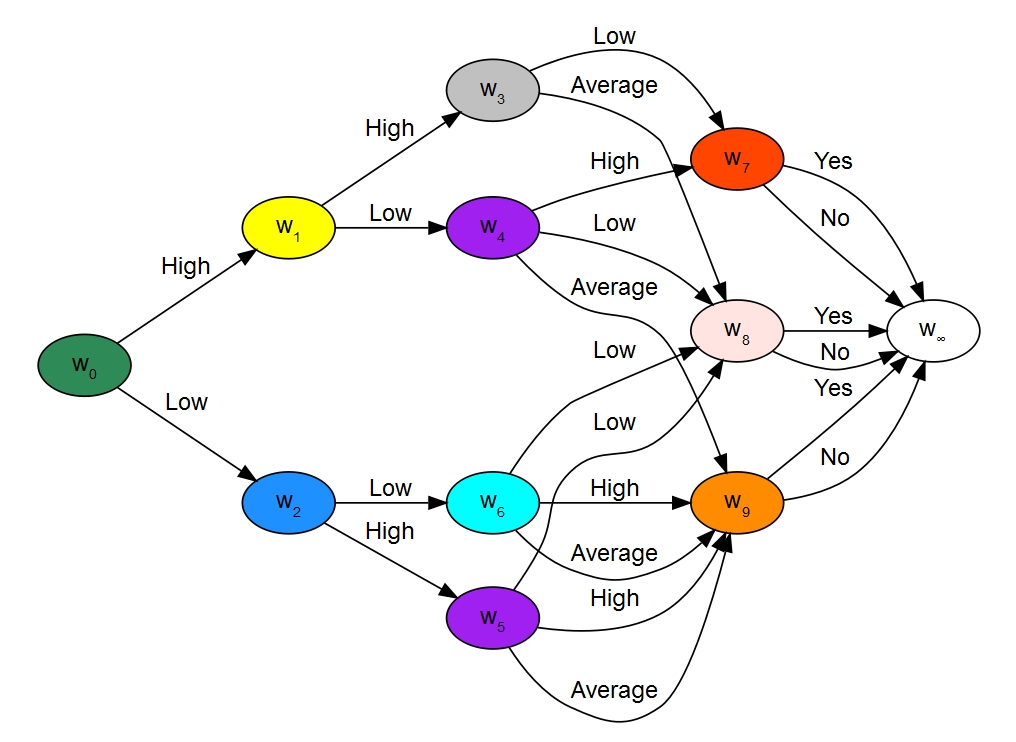} 
		\caption{CEG$_\text{BN}$, a CEG adapted from the BN used in previous CHDS study.}
		\label{fig:chdscegBN}
	\end{figure}
	
	\begin{figure}[h] 
		\centering
		\includegraphics[width=4.5in]{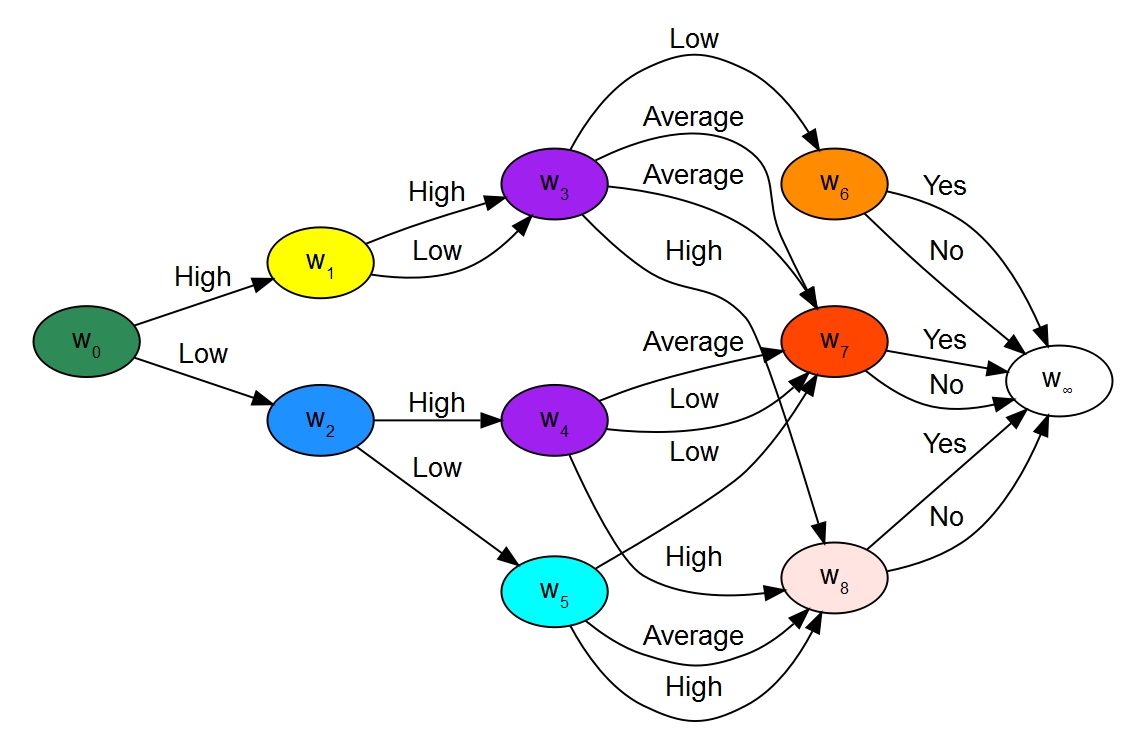} 
		\caption{CEG$_\text{AHC}$, The CEG for the CHDS data found using the AHC algorithm.}
		\label{fig:chdscegB}
	\end{figure}
	\subsection{Event trees}
	
	A CEG is built from a coloured event trees and its construction is well explained elsewhere \citep{CEGbook}. Here we briefly review this construction. We first introduce event trees, then a colouring, then a class of staged trees, then a much simpler graph derived from the staged tree--a CEG. This formal development leads us through increasingly fine features of the CEG. These will correspond to the relevant diagnostic monitors we develop later.
	
	Let $\mathcal{T} =(V,E)$ denote a directed tree with $V$ and $E$ denoting the node and edge set respectively. 
	The set of vertices $\text{pa}(v) = \{v' \,|\, \text{there is } (v',v) \in E\}$ represents the parents of $v \in V$ 
	and $\text{ch}(v) = \{v' \,|\, \text{there is } (v,v') \in E\}$ denotes the children of $v \in V$.
	It is often helpful to distinguish between the vertices which are situations $s \in S$ and the leaf nodes $l \in V \setminus S$. Situations nodes are non-leaf nodes in the event tree. 
	We can denote the set of root-to-leaf paths in an event tree by $\Lambda(\mathcal{T})$. $\Lambda(v)$ and $\Lambda(e)$ refer to vertex-centered and edge-centered events, the subset of root-to-leaf paths that pass through either the vertex $v$ or edge $e$.
	For a particular situation $v \in V$ and its emanating edges $E(v) =\{(v,v') \in E | v' \in \text{ch}(v)\}$, we can define a floret as the pair $\mathcal{F}(v) = (v,E(v))$.

	We next assign a probability distribution to this event tree with parameters $\theta(e) = \theta(v,v')$ corresponding to the edge $e = (v,v') \in E$. The components of all floret parameter vectors sum to unity $\sum_{e \in E(v)} \theta(e) = 1$ for all $e \in E$ and $v \in V$. 
	Each parameter $\theta(e), e \in E$ is a primitive probability. These primitive probabilities serve a similar role to potentials in BNs.
	The pair $(\mathcal{T},\bm{\theta}_{\mathcal{T}})$ of a graph $\mathcal{T}$ and all labels $\bm{\theta_{\mathcal{T}}} = (\theta(e) | e \in E)$ is called a probability tree.
	
	
	\subsection{Staged trees}
	
	Building on the definition of a probability tree as the pair $(\mathcal{T}, \bm{\theta}_{\mathcal{T}})$ with graph $\mathcal{T}=(V,E)$ and labels $\bm{\theta_{\mathcal{T}}} = (\theta(e) | e \in E)$, we can now define a staged tree. The stagings represent conditional independence in the CEG. 
	Two vertices representing situations $v, v' \in S$ are in the same stage $u$ if and only if their floret distributions are equal up to a permutation of their components $\bm{\theta}_v = \bm{\theta}_{v'}$.
	Each stage is assigned a unique colour. 
	An event tree can be transformed to a staged tree by colouring the vertices according to their stage memberships. 
	If all vertices are either in the same stage or have pairwise different labels, then $(\mathcal{T}, \bm{\theta}_{\mathcal{T}})$ is a staged tree.
	
	This results in a set of stages of the staged tree denoted as $U$, defined as:
	
	\begin{equation}
	U = \{u \subseteq V \,|\, v \text{ and } v' \text{ are in the same stage for all } v,v' \in u \}.
	\end{equation}
	
	There is a finer partition of events called positions $w \in W$. We denote $\mathcal{T}(v) \subseteq \mathcal{T}$ as the event tree rooted at $v \in V$ and whose root-to-leaf paths are inherited from $\mathcal{T}$. Then we can say that the pair $(\mathcal{T}(v), \bm{\theta}_{\mathcal{T}(v)})$ is a probability subtree of $(\mathcal{T}(v), \bm{\theta}_{\mathcal{T}(v)})$.  Two situations $v, v' \in u$  which are in the same stage $u \in U_{\mathcal{T}}$ are also in the same position if their subtrees $(\mathcal{T}(v), \bm{\theta}_{\mathcal{T}(v)})$ and $(\mathcal{T}(v'), \bm{\theta}_{\mathcal{T}(v')})$ have the same graph and the same set of edge labels.
	Visuals of the event trees and subsequent staged trees for the CHDS example can be found in \citet{CEGbook}.
	
	\subsection{Chain Event Graphs}
	Building on the concepts of stages and positions, a CEG can be constructed from a staged event tree by merging situations that lie in the same position. Formally, a CEG $\mathcal{C}(\mathcal{T})=(W,F)$ is the pair of positions $W$ and accompanying edge set $F$. The vertex set $W = W_{\mathcal{T}}$ is the set of positions in the underlying tree $\mathcal{T}$. Each position $w$ inherits its colour $u$ from the staged tree. 
	If all edges  $e = (v_1, v_2'),\, e' = (v_2,v_2') \in E$ and the vertices $v_1,v_2$ are in the same position, then there is a corresponding edge $f,f' \in F$. The labels $\theta(f)$ of edges $f \in F$ are inherited from the corresponding edges in the staged tree. 
	The labelled graph
	$(\mathcal{C}(\mathcal{T}),\bm{\theta}_{\mathcal{T}})$ is a Chain Event Graph.

	The partition specifying the stages a CEG is analogous to specifying conditional independence asserted through the graph of a BN \citep{Dawid1979,Studeny2002}. Situations in the same stage are independent conditional on their respective histories and the proofs of  can be found in \citet{SmithAnderson2008, Thwaites2010}.
	
	For this paper, we consider the class of stratified CEGs because they offer the most direct comparison to a standard BN. 
	In the class of stratified CEGs, the atoms (i.e. the root-to-sink paths) of every CEG are identified with elements in the product state space of the ordered set of random variables $\bm{X} = (X_1, \ldots, X_i, \ldots,X_n)$ where every component $X_i$ has a set number of levels, $K_i$, such that each of the levels is the same distance from the root node.

	Current search algorithms have been developed for stratified CEGs that search the space of trees.  
	These include dynamic programming methods \citep{Cowell2014} and an Agglomerative Hierarchical Clustering algorithm \citep{Freeman2011}. A greedy search algorithm may miss the optimal model, further reason to check the model using our diagnostics.
	Further adaptations of these search methods have been developed including a search method based on Bayesian Information Criterion (BIC) \citep{Schwarz1978}. 
	These algorithms have been implemented in \citep{stagedtrees}.
	Search for asymmetric structures is currently being developed, as are extensions to search over a range of variable orderings \citep{CEGbook}.
	

	
	The CEG$_\text{BN}$ in Figure \ref*{fig:chdscegBN} encodes the same conditional independence relationships as the BN in Figure \ref{fig:chdsBN}.
	The BN in Figure \ref{fig:chdsBN} models that $X_{\text{h}}$ is independent of $X_{\text{e}}$ given $X_{\text{l}}$ and $X_{\text{s}}$. $\text{CEG}_{\text{BN}}$ in Figure \ref{fig:chdscegBN} encodes this through the colouring in the set of stages representing $X_{\text{h}}$. 
	For $X_{\text{s}}=$ High (or Low), the future development of $X_{\text{l}}$ is not dependent on $X_{\text{e}}$
	The edges for both levels of $X_{\text{e}}$ go into the same stages. $\text{CEG}_{\text{AHC}}$ in Figure \ref{fig:chdscegB} represents the CEG found by the AHC algorithm.
	
	\section{Prequential diagnostics for a BN}\label{sec:diag}
	
	\subsection{Conjugate Dirichlet analysis in the BN}
	
	A Bayesian Network $G$ is given by a set of random variables $X_i$ for $\{i \in 1, \ldots, n\}$, each taking different values  $x_k$ for $\{k \in 1, \ldots, K_i \}$. The possible configurations of the parents of $X_i$ are denoted $\rho_i= j$ are $\{ 1 , \ldots, q_i\}$.   
	We can set a Dirichlet prior for each set of parents of node and values of $X_i$ governed by parameter $\bm{\theta}_{ijk}$. 
	
	Suppose we observe $\bm{y}_i = \{y_1, \ldots, y_m , \ldots, y_M\}$, a series of observations for the variable $X_i$, where each possible value of each random variable is assigned a Dirichlet prior $\mathcal{D} (\alpha_1, \ldots, \alpha_K)$.
	In a discrete BN, the entries in the conditional probability tables for a particular parent setting sum to one over all possible levels of the node. That is, the parameter for the $i$th node with the $j$th setting of the parents for the $k$th value,  $\theta_{ij} =\sum_{k=1}^{K_i} \theta_{ijk} =1$. 
	We can set a Dirichlet prior for each $\theta_{ij}$, and use the conjugate posterior analysis.
	As data is accumulated about the system, the Dirichlet prior can be updated by adding the counts of the observation to the prior. 
	We can compute a reference Dirichlet prior by taking the highest number of levels of a given variable ($X_{\text{l}}$ gives an effective sample size of $\alpha = 3$ for the CHDS example) and dividing it by the number of levels outgoing from each situation.

	The prequential diagnostics compute the surprise of seeing each subsequent observation given the past observations. Towards that end, our monitors use the likelihood of observing the complete data $\bm{y}$ as given by \cite{Heckerman1995}. Assuming it was randomly sampled, the likelihood of the probability vectors is: 
	\[
	p(\bm{y} | \bm{\theta}) = \prod_{i=1}^{n} \prod_{j=1}^{q_i} \prod_{k=1}^{K_i} \theta_{ijk}^{y_{ijk}}
	\]
	
	The parameter for each value and parent pair for each node $\theta_{ijk}$ is governed by a Dirichlet distribution. Thus the prior is given by: 
	
	\[
	p(\bm{\theta})= \prod_{i=1}^{n} \prod_{j=1}^{q_i} \frac{\Gamma(\sum_{k=1}^{K_i} \alpha_{ijk})}{\prod_{k=1}^{K_i} \Gamma(\alpha_{ijk})} \prod_{k=1}^{K_i} \theta_{ijk}^{\alpha_{ijk}-1}
	\]
	
	Following the conjugate analysis, we obtain the following form of the marginal likelihood: 
	

	%
	
	\begin{equation}
	p(\bm{y}) = \prod_{i=1}^{n} \prod_{j=1}^{q_i} \frac{\Gamma(\sum_{k=1}^{K_i}\alpha_{ijk})}{\Gamma(\sum_{k=1}^{K_i}\alpha_{ijk+})} \prod_{k=1}^{K_i} \frac{\Gamma(\alpha_{ijk+})}{\Gamma(\alpha_{ijk})}
	\end{equation}
	where $\alpha_{ijk+} = \alpha_{ijk} + y_{ijk}$.
	%
	
	\subsection{Scoring rules}
	
	In order to check the accuracy of the forecasts, we can use the logarithmic scoring rule.
	
	
	Let $y_m$ denote the $m$th observation of the data for which $y_m$ is observed at a specific level of the random variable $y_k$.
	$p_m$ is the predictive density of observing $y_k$ after learning from the first $m-1$ cases. 
	The logarithmic score of the $m$th observation of $Y$ taking the value $y_k$ is denoted:
	\[
	S_m = -\log p_m(y_k)
	\]
	
	There are two methods of standardisation. Relative standardisation examines the logarithmic difference between the penalties under two different models. 
	The absolute difference does not require an alternative model. 
	Instead, we compute a standardised test statistics $Z_m$ using the expectation $E_m$ and variance $V_m$ following \citet{diagnostics}: 
	
	\begin{align}\label{eq:standard}
	E_m &= -\sum_{k=1}^{K} p_m(y_k) \log p_m(y_k)\\
	V_m &= \sum_{k=1}^{K} p_m(y_k) \log^2 p_m(y_k) - E_m^2\\
	Z_m &= \frac{\sum_{m=1}^{M}S_m - \sum_{m=1}^{M}E_m}{\sqrt{\sum_{m=1}^{M}V_m}}
	\end{align}
	
	For sufficiently large sample sizes under the model assumptions, $Z_m$ will have a standard Normal distribution if the model could have plausibly generated the data. 

	\begin{figure}[h] 
		\centering
		\includegraphics[width=2in]{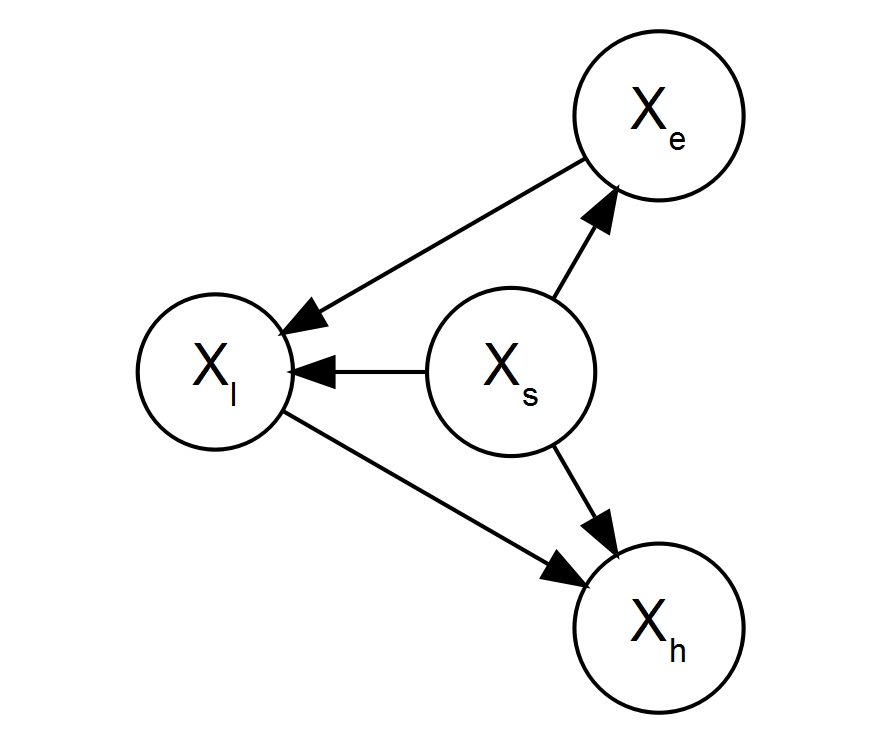} 
		\caption{BN CHDS: A BN obtained from previous studies of the CHDS data \citep{Barclay2013}.}
		\label{fig:chdsBN}
	\end{figure}
	
	For the global monitors, we can now examine alternative models under the relative standardisation technique. 
	Our candidate models include the baseline BN shown in Figure \ref{fig:chdsBN}, a CEG based on the BN that includes additional information in Figure~\ref{fig:chdscegBN}, and another CEG found from the AHC algorithm in Figure~\ref{fig:chdscegB}. 
	This enables us to identify structural improvements with an increasingly fine set of monitors. 
	
	\subsection{Diagnostic monitors for Bayesian Networks}
	
	The prequential methods are similar to cross-validation, with the key difference being that they rely on information from the previous iterations, rather than predicting on the variables excepting the one of interest.

	Within a Bayesian framework these diagnostics are especially attractive, because if the estimated conditionals are treated as one-step ahead predictives, then the log marginal likelihood is simply the sum of these scores. So the prequential methods then decompose an aggregate score into scores associated with different subsets of the contributions to the data. Each such subset can then be scrutinized for its fidelity to the fitted model as it applies to that subset within the context of a full Bayesian analysis. 
	
	When most effective, the prequential approach is able to adopt an interpretable and natural ordering of the observational data.
	When a temporal component is not immediately obvious, it may be helpful to order the data according to some covariate of the observables. For instance, modelling healthcare outcomes might benefit from ordering the data according to the length of time each observation spent in the hospital. 
	The prequential approach is well suited to detect where the model is no longer a good fit to the data.
	
	The monitors discussed in \citet{diagnostics} that we reproduce for the BN include the global monitor for overall model fit, the node monitor to check the probability distributions, and the parent-child monitor to assess the contribution of individual parent settings. 
	
	\paragraph{Global monitors}\label{sec:bnGlobal}
	
	The global monitor for BNs is defined as the logarithmic probability of the $m$th observation 
	:
	$-\log p_m (y_m)$
	after $m-1$ cases are processed.
	The overall global monitor for all $M$ cases is:
	\begin{align}
	G_{\text{BN}} &= -\log\prod_{m=1}^{M} p_m(y_m) =-\log\prod_{m=1}^{M} p_m(y_m | y_1, \ldots, y_{m-1}) \label{eq:e}\\
	&= -\log p(y_1 \ldots, y_m)= -\log p(\bm{y}).\label{eq:z}
	\end{align}

	Calculating the global monitor for two different systems provides an immediately interpretable comparison between models. 
	These monitors have been shown to provide quick checks of BN structure against data. 
	To illustrate, the log marginal likelihood, equivalent to the global monitor, for BN CHDS is $G_{\text{BN}} =-2495.01$. In Section~\ref{sec:ex}, we will see how this compares to the global monitor of competing models.

	\paragraph{Node monitors}
	
	The node monitor assesses the adequacy of the marginal and conditional probability distributions for each node in the model. 
	The marginal node monitor is given by 
	\[
	N_{\text{marg}} = -\log p_m(x_k)
	\]
	
	\noindent after $m-1$ cases are processed. This is calculated by ignoring the other evidence in the $m$th case after $X_i$ is observed. The unconditional node monitor checks the suitability of the probability distribution of the node. 
	
	The conditional node monitor use probabilities that are conditioned on evidence in the $m$th case. To compute the conditional node monitor, all of the evidence in $\mathcal{E}$ is propagated except for $X_i = x_i$. Then the conditional node monitor can be represented as: 
	
	\[
	N_{\text{cond}} = -\log p_m(x_i | \mathcal{E}_m \setminus X_i).
	\]
	
	The conditional node monitor checks how well the model predicts each node given the other evidence in the observation. First, we specified the conditional probability tables, with $\theta_i$ after learning from the first $m-1$ cases.
	For the conditional node monitors, we propagated the evidence from the other variables omitting the node under consideration, and then queried the BN with the functions in the R package \texttt{gRain}~\citep{grain}. 
	
	For instance, to compute the conditional node monitor for $X_\text{h}$, we propagated the evidence $\mathcal{E} = \{ X_\text{s} = \text{High}, X_\text{e} = \text{High}, X_\text{l} = \text{High}, X_\text{h} = \text{Yes} \}$ and queried $X_\text{h}$ according to the structure in Figure~\ref{fig:chdsBN}. 
	The node monitors are then standardised according to Equation \ref{eq:z}.
	\begin{table}
		\centering
		\begin{tabular}{crr}
			$X_M$ & $Z_{\text{marg node}}$ & $Z_{\text{cond node}}$\\
			\hline \\
			$X_\text{s}$  & 1.708 & 0.1737 \\
			$X_\text{e}$  & 0.582 & -1.560 \\
			$X_\text{l}$  & 2.953 & 2.454 \\ 
			$X_\text{h}$  & 0.340 & -0.450 \\ 
		\end{tabular}
		\caption{Final BN node monitors for the CHDS example}
		\label{tab:bnNode}
	\end{table}
	
\begin{figure}
		\centering
		\begin{subfigure}[b]{0.47\textwidth}
			\centering
			\includegraphics[width=\textwidth]{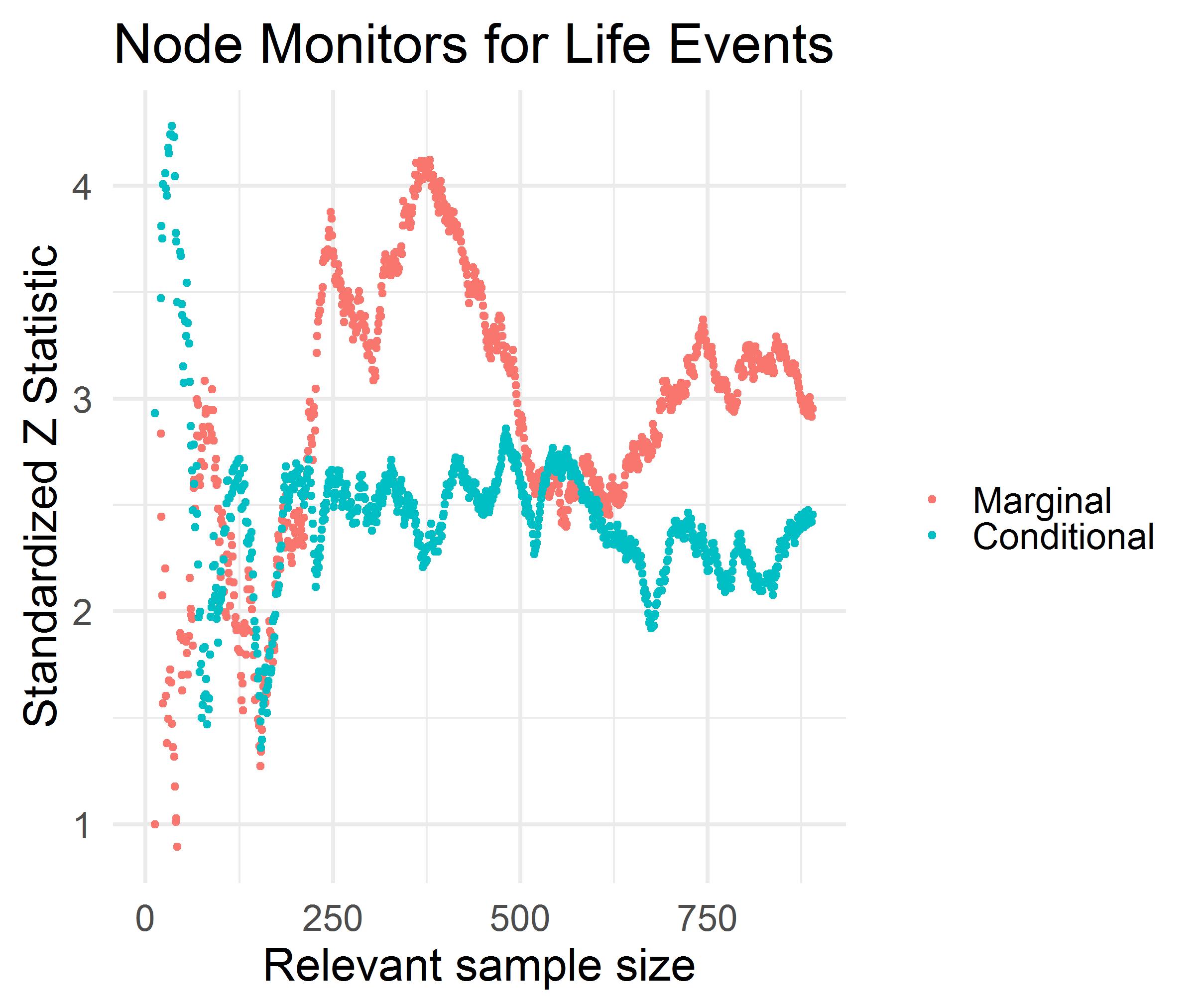} 
			\caption{The marginal and conditional node monitor for $X_\text{l}$. 
			}
			\label{fig:nodeBNevents}
		\end{subfigure}
		\hfill
		\begin{subfigure}[b]{0.47\textwidth}
			\centering
			\includegraphics[width=\textwidth]{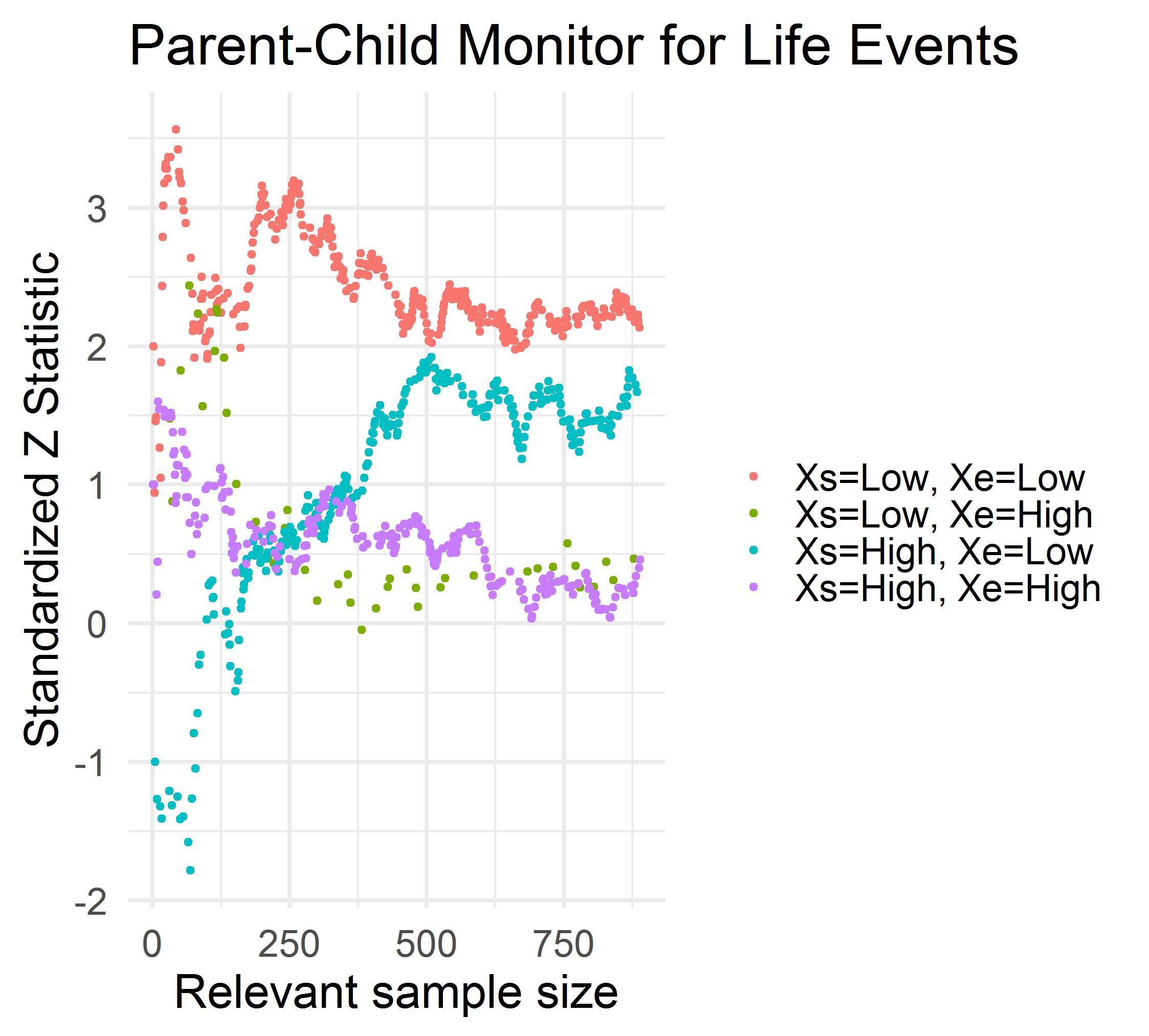}
			\caption{CHDS BN: the parent-child monitor for positions all possible parent settings of $X_\text{l}$} 
			\label{fig:pachBNevents}
		\end{subfigure}
		\caption{Node monitors detect ill-fitting distribution for $X_{\text{l}}$}
	\end{figure}

	Computing the final node monitors offers a quick check to see which node probability distributions might be incorrectly specified. The final node monitors for the CHDS BN are shown in Table \ref{tab:bnNode}. The marginal and conditional node monitors for $X_\text{s}$, $X_\text{e}$, and $X_\text{h}$ are properly calibrated. However, we notice that the predictive probability distribution appears to be misspecified for $X_\text{l}$. The plot in Figure~\ref{fig:nodeBNevents} confirms that both the marginal and conditional monitors indicate that we should not trust the modelling of $X_\text{l}$.

	As we will see in Section \ref{sec:cegdiag}, the nodes of the BN are not exactly analogous to the positions of a CEG. Additional checks on the stages and the situations composing the stages will be required.
	
	\paragraph{Parent-child monitors}
	
	After identifying the problematic node, the parent-child monitor can be used to pinpoint the configurations of the parent values which might be associated with the misspecification.
	For any node $X_i$ in a BN (noting that this is distinct from the situations and vertices $v$ in a CEG), the parent-child monitor is defined as the predictive posterior of the $m$th observation with parents $\rho$ after learning from the first $m-1$ cases with parents $\rho$:

	\[
	R = p_m(x_i| X_{\text{pa}(i)}^{m-1}=\rho).
	\] 
	
	Historically, the parent-child monitor has been used to confirm the effects of learning and the selected priors on the model  \citep{diagnostics}. 
	The parent-child monitor can also be used to assess the appropriateness of different priors on individual nodes. We use it here to identify BNs that have context-specific probability distributions that could be remedied by reexpressing the problem as a CEG.
	A good heuristic is that any predictive model with $|Z_m| > 1.96$ should be viewed with suspicion \citep{Cowell2007}.
	
	
	In Figure~\ref{fig:pachBNevents}, we check the parent-child monitor for $X_l$ given all possible parent settings. This indicates that the household with $X_\text{s}=$ Low and $X_\text{e}=$ Low are a particularly poor fit to the data.
	Because the parent-child monitor assesses how sensitive a model is to particular setting, we use it here to indicate when a BN should be adapted to a CEG model.

	\section{Diagnostics for a CEG}\label{sec:cegdiag}
	
	The monitors below explain what we might expect to see from the model in a predictive space. Prequential monitors can pinpoint where and how forecasts from candidate models deviate. 
	The model fit might deviate because there can be two different data generating processes, and in this situation we might want to use the diagnostics to help explain why one model is a better fit than another. Additionally, data exchangeability might not hold, or the data might have some other built up dependence that the current structure does not capture.  
	
	\subsection{Conjugate Dirichlet analysis in the CEG}
	Within a conjugate analysis, product Dirichlet-Multinomial distributions describe the posterior and more importantly the predictive distributions we use in our specific prequential analysis.
	Suppose we have either elicited or used model selection techniques to acquire the CEG, $\mathcal{C}$ with $K$ stages denoted $u_1, \ldots, u_K$.
	Each stage $u_i$ in $\mathcal{C}$ has floret parameters $\bm{\theta}_i$ for $i \in 1, \ldots, K$.
	Edges in a stage are $E(u_{i}) = \{e_{i1}, \ldots, e_{iK_i}\}$ with labels $\theta_{ij} = \theta(e_{ij})$ for $j = 1, \ldots, K_i$ and $i =1,\ldots, K$. 
	Then suppose we observe a sample $\bm{Y} = \bm{y}$. From this we know in part how many observed counts arrive at each of the $K$ stages. We denote the counts at each individual stage as $\bm{y} = (\bm{y}_0, \ldots, \bm{y}_i, \ldots \bm{y}_K)$ where $\bm{y}_i = (y_{i1},\ldots, y_{ij}, \ldots, y_{iK_i})$. 
	
	Assuming that the experiment was randomly sampled, then the floret parameter vector $\bm{\theta}_i$ has a Multinomial distribution $\text{Multi}(N_i, \bm{\theta}_i)$ where $N_i = \sum_{j=1}^{K_i} y_{ij}$ whose mass function we denote as $p_i(\bm{y}_i | \bm{\theta}_i)$. The separable form of the likelihood of the probability vectors for stages $u_1, \ldots, u_K$ is given by: 
	
	\[
	p(\bm{y}| \bm{\theta}) = \prod_{i=1}^{K} p_i(\bm{y}_i | \bm{\theta}_i) = \prod_{i=1}^{K} \prod_{j=1}^{K_i} \theta_{ij}^{y_{ij}}.
	\] 
	
	The Dirichlet prior distribution for each of the stages is denoted as $\bm{\alpha}_i = (\alpha_{i1}, \ldots, \alpha_{iK_i})$. Thus the prior is given by: 
	
	\[
	p(\bm{\theta}) = \prod_{i=1}^{K} \frac{\Gamma(\sum_{j=1}^{K_i} \alpha_{ij})}{\prod_{j=1}^{K_i} \Gamma(\alpha_{ij})} \prod_{j=1}^{K_i} \theta_{ij}^{\alpha_{ij}-1}.
	\]
	
	Following the conjugate analysis in \citet{CEGbook}, under closed sampling we obtain the following form for the marginal log likelihood:

	%
	%
	%
	%
	
	\begin{align}\label{eq:cegLoglike}
	\log p(\bm{y}) = \sum_{i=1}^K  \left[\log \Gamma(\bar{\alpha}_i) - \log \Gamma(\bar{\alpha}_{i+}) - \left(\sum_{j=1}^{K_i} \log \Gamma(\alpha_{ij}) - \log \Gamma(\alpha_{ij+}) \right) \right]
	\end{align}
	
	\noindent where $\bar{\alpha}_i = \sum_{j=1}^{K} \alpha_{ij}$ for all $i \in 1, \ldots, K$ and $\alpha_{i+}= \alpha_i + y_i$.
	
	
	\subsection{Global monitor}
	
	As shown in Section~\ref{sec:bnGlobal}, the global monitor is the probability of observing all of the evidence for a particular case $m$ after processing $m-1$ cases, $P_m(\mathcal{E}_m)$. Evidence for the CEG is defined as the root-to-leaf path containing the observation. The overall global monitor then is defined as the product of observing each of the $m$ cases: 
	\[
	G_{\text{CEG}} = -\log p(y_1, \ldots, y_m) =  -\log p(\bm{y})
	\] 
	
	For a CEG, this is given by the marginal likelihood $p(\bm{y})$ shown in Equation \ref{eq:cegLoglike}.
	The global monitor offers an immediately interpretable comparison of candidate models. It also defines a way to directly compare a CEG equivalent to a BN with a CEG found using another method, as we see for the CHDS example in Section \ref{sec:ex}. After making changes to finer aspects of the structure, the global monitor may be computed to show improvements in the overall model.
	
	\subsection{Staging monitors}
	
	Staging monitors are designed to identify problems with the staging of the colourings for a given cut (a variable in the stratified CEG). This does not have an analogy to the BN monitors because it is designed to detect discrepancies within the context-specific conditional independence relationships and ordinary BNs do not accommodate such structure. However, it can be used on a CEG representation equivalent to a BN to detect particular context-specific independences within this class.
	
	Staging monitors are the staging predictive distributions, $p(U_t | \bm{y}^{m-1})$. 
	$U_m$ denotes the partitioning of stages $u_1, \ldots u_K$ in a particular cut and $\bm{U}$ represents all the possible alternative partitions. 
	The form of the one step ahead predictive allowing for first-order Markov transitions between stages is given in \citet{Freeman2011a}. 
	Because our primary aim is to see if the model staging is an appropriate fit given the data, we do not allow for transitions between stagings. 
	We consider the set of alternative stagings to be the stagings that are one move on the Hasse diagram away from the given staging. 
	
	
	
	%
	
	To assess the appropriateness of the staging to the data, we need the quantity: 
	
	\[
	p(U_{m} = U' | \bm{y}^{m-1}) \propto p(\bm{y}_{m-1} | U_{m-1} = U')p( U_{m-1} = U' | \bm{y}^{m-2})
	\]
	
	\[
	= \frac{p(\bm{y}_{m-1} | U_{m-1} = U')p( U_{m-1} = U' | \bm{y}^{m-2})}{\sum_{U' \in \bm{U}} p(\bm{y}_{m-1} | U_{m-1} = U')p( U_{m-1} = U' | \bm{y}^{m-2})}.
	\]
	
	As shown in \citet{Freeman2011a}, $P( U_{m-1} = U' | \bm{y}^{m-2})$ is available at time $t-1$ and 
	
	\[
	p(\bm{y}_{m-1} | U_{m-1} = U') = \int_{\bm{\Theta}_{m-1}} p(\bm{y}_{m-1} | \bm{\theta}_{m-1}, U_{m-1} = U') p (\bm{\theta}_{m-1} | U_{m-1} = U') d\bm{\theta}_{m-1}
	\]
	
	\[
	=\prod_{i=1}^K \frac{\Gamma(\sum_{j=1}^{K_i} \alpha_{m-1}^{ij})}{\Gamma(\sum_{j=1}^{K_i} \alpha_{m-1}^{ij+})} \prod_{j=1}^{K_i} \frac{\Gamma(\alpha_{m-1}^{ij+})}{\Gamma(\alpha_{m-1}^{ij})}.
	\]

	Here we have embedded the time index so that $\alpha^{ij}_{m-1}$ denote $\alpha_{ij}$ at time $t-1$ and $\alpha^{ij+}_{m-1}$ denote $\alpha_{ij} + y_{ij}$ at time $t-1$.
	The staging monitor identifies places where the data is no longer a good fit for the existing stage structure. 
	
	The plots of the staging monitor depict $p(U_m = U' | \bm{y}^{m-1})$ for the assumed stage $U$ and each alternative staging $U'$ over the number of observations in the dataset. 
	This allows us to see how the suitability of the model changes over time. 
	If one of the alternative stages emerges as the highest probability forecast, then this indicates that the alternative staging in the model class should be used instead. 
	If no clear staging emerges, this indicates that the appropriate staging may be outside the model class. This could indicate that the data-generating process draws from different stagings at different times.
	
	We will see how this enables us to differentiate between possible stagings in the CHDS data in Section~\ref{sec:ex}.
	
	\subsection{Position monitors}
	
	The node monitors for a BN detect discrepancies in the probability distribution specified for each node. For the CEG, we want to check the probability distributions specified for each position. 
	Mirroring the BN methodology, we will compute a marginal and conditional probability. 
	
	To compute the marginal position monitor, $N_{\text{marg}}$ for the $m$th observation in our dataset, we first compute the probability florets for each of the positions based on the previous $m-1$ observations in the dataset.
	Because positions only apply to data that matches the appropriate upstream pathways of $w_i$, the position monitors are only computed for those observations. 
	Then, we want to know what the marginal probability of observing the $m$th observation take each of the values $e_k$ emerging from the position $w_i$. We compute these by summing the probability of each of the root to sink paths that goes through the edge of interest $e_k$. The marginal monitor is given by: 
	\[
	N_{\text{marg}} = -\log p_m (\Lambda(\theta(w_i) = e_k))
	\]
	The marginal monitor is then standardized against the actual observed value of $w_i$ in the $m$th observation according to the Equations~\ref{eq:standard}.
	
	
	The conditional node monitor computes the probability of observing evidence for the $m$th case after propagating evidence from the observations in the $m$th observation, excluding the outcome in the position of interest $w_i$. 
	The conditional node monitor was designed for BNs to check the appropriateness of a distribution for a node conditional on the evidence for all the other nodes in the BN.  
	As the CEG is automatically conditioning on all of the upstream variables, the conditional monitors for the positions of a CEG only provides information additional to the marginal node monitor for certain structures defined below. 
	Like the marginal position monitor above, these are functions of observations within a given position of interest.
	
	Whereas with the marginal monitor, we can compute the marginals from the probability florets directly, we need to use message passing to pass the evidence to update the probability florets for the conditional monitors. 
	The propagation algorithm for the CEG is given in \citet{ThwaitesCowell2008} with additional details in \citet{CEGbook}.
	The propagation algorithm relies on evidence, which is the full root-to-sink path in the CEG. 
	The evidence for the $m$th observation is some subset of the settings of random variables at the $m$th observation.
	
	Evidence is propagated through a sub-graph of the CEG called the transporter. 
	The transporter inherits the probabilities $\theta(w_i)$ for the set of positions and edges in the transporter. In the conditional monitor, we compute the $p_m$ from the probabilities from the previous $m-1$ cases. 
	The probabilities are back-propagated, i.e. summed at each position to compute the potential, $\phi(w_i)$. 
	Then the probabilities are then updated by dividing each $\theta(w_i)$ by $\phi(w_i)$. 
	Thus, if the potential for $w_i$ sums to one, then the updated probabilities are the same as the original.
	The conditional position monitor is given by: 
	\[
	N_{\text{cond}} = -\log p_m(\theta(w_i) = e_k | \mathcal{E}_m \setminus \theta(w_i)).
	\]

	\noindent 
	The conditional monitors are then standardized according to Equation~\ref{eq:standard}. 
	For our examples in Section~\ref{sec:ex}, $\phi(w_i) = 1$, so for our example, we need show only the marginal monitors. 
	
	The position monitors can be compared to a BN node monitor to confirm the suitability of the CEG structure. Within the CEG model class, it can detect discrepancies within the specified probability distribution. If the marginal position monitor indicates a poor fit, but the conditional position monitor indicates an appropriate fit, then we may continue cautiously using the selected model. However, if both the marginal and conditional position monitors indicate a poor fit, then we may want to consider alternative models. 
	The monitors are designed to be used from the coarsest to finest, so we would only detect an issue with the position after confirming that the staging is appropriate. Thus the position monitor detects issues that may be at or downstream of the position. Perhaps certain situations that are in the same stage should not be in the same position. 
	The position monitor can also be used to detect when data has been generated from a model with additional positions or information available.
	

	
	\subsection{Situation monitors}
	
	At the finest level, the CEG is composed of situations defined in Section~\ref{sec:ceg}. 
	A stage $u_i$ in a CEG is composed of situations $\{v_1, \ldots, v_k, \ldots,  v_M\}$ that are by definition exchangeable. 
	A situation monitor highlights situations when this exchangeability assumption might be violated. 
	
	The prequential methods check the validity of the forecasts. To check the forecasts from each of the stages in the structure, we need to compare the forecasts coming from each of the different situations. The stage order monitor imposes a new order to retain the prequential methodology. 
	The leave one out monitor stage monitor offers a quick check and additional aid to model transparency.

	\paragraph{Leave one out stage monitor}
	
	Using a method similar to the leave one out cross validation, we can examine the Bayes factor contribution from the stage $u_i$ with a particular situation $v_k'$  removed, denoted $f(\bm{y}_{i,-k}')$, and compare it to the Bayes factor contribution from the stage as a whole, $f(\bm{y}_{i})$ as above. We expect that the stage with all contributing situations to be preferable to the one with the situation removed. Thus, this offers a quick check if any removing any situations leads to a higher Bayes factor score. We refer to this as the leave one out monitor, given by 
	%
	
	\[
	Q(u_k, v_k) = \log f(\bm{y}_{i,-k'}) -\log f(\bm{y}_{i})
	\]
	where the contribution from the stage with situation $v_k$ left out is
	\[
	\log f(\bm{y}_{i,-k'}) = \log \Gamma(\bar{\alpha}_{i}) - \log \Gamma(\bar{\alpha}_{i+,-k'}) - \left(\sum_{j=1}^{K_i} \log \Gamma(\alpha_{ij}) - \log \Gamma(\alpha_{ij+,-k}) \right)
	\]
	
	where $\alpha_{i+,-k'} = \alpha_i + y_{i,-k}$.
	A quick visual check can plot the actual observed proportions in each situation against the proportion we expect to see from the predictive posterior with data from the stage of interest missing. We examine the proportions of a particular level $l = l'$ for each of the stages. The stages associated with the variables that take extreme values are often of particular interest. For instance, for $X_{\text{h}}$ in the CHDS data, we consider the proportion of households for which $X_{\text{h}}=$ Yes. 
	
	We could use this for more than two levels, but it would be more difficult to picture the discrepancy, and thus more difficult to display and interpret the output. 
	Reducing the problem to a binary question allows us to leverage the  properties of the Dirichlet distribution closure to marginalisation \citep{CEGbook}.
	We can compute the conjugate posterior $\text{Beta} (\alpha', \beta') = (\alpha^+_{-k'}, \beta^+_{-k'})$ with the situation $v_{-k}$ removed and take the expectation $\frac{n\alpha'}{(\alpha' + \beta')}$ where $\alpha'$ corresponds to the level of interest $l=l'$. We can compare this to the observed proportion of units where  $y_i = l'$. 
	
	\paragraph{Situation order monitor}
	
	To use a prequential check on the stage structure, we can impose an ordering on the relevant situations $\{v_1, \ldots, v_{\tilde{m}}, \ldots, v_{\tilde{M}}\}$. This ordering could correspond to some notion of severity of the situations. For instance, in the CHDS data, we might order the situations in cut $X_{\text{l}}$ according to increasing adversity $I(X_{\text{l}}) = $ Low, Average, High. Imposing this ordering ensures that the corresponding residuals are independent.
	
	We can then compute the predictive distribution after observing the distribution of data from the first $\tilde{m}-1$ situations $v_{1 <\tilde{m} < \tilde{m} -1}$ using the posterior with distribution $\text{Beta} (\alpha_{\prec \tilde{m}}, \beta_{\prec \tilde{m}})$  where $\alpha_{\prec \tilde{m}} = \alpha_i + \sum_{\tilde{m}=1}^{\tilde{m}-1} y_{i\tilde{m}}$ and $\beta_{\prec \tilde{m}}= \beta_i + \sum_{\tilde{m}=1}^{\tilde{m}-1} y_{i\tilde{m}}$ represent the count data from only the preceding situations.
	The surprise of observing the number of counts $y_{\tilde{m}'l'} $  of the `worst' level in the subsequent situation $v_{\tilde{m}'}$ is given by:
	
	\[
	p_{\text{order}} = -\log p_{\tilde{m}} (\bm{y}_{\tilde{m}}^i = y_{\tilde{m}}^{i,k} \,|\, \bm{y}_{\tilde{m}-1})
	\]
	
	\[
	p(y_{\tilde{m}'l'}) = \frac{\Gamma(\alpha_{\prec \tilde{m}'} + \beta_{\prec \tilde{m}'})}{\Gamma(\alpha_{\prec \tilde{m}'})\Gamma(\beta_{\prec \tilde{m}'})} \binom{y_{\tilde{m}'}}{y_{\tilde{m}'l'}} \frac{\Gamma(\alpha_{\prec \tilde{m}'} + y_{\tilde{m}'l'}) \Gamma(\beta_{\prec \tilde{m}'} + y_{\tilde{m}'} -y_{\tilde{m}'l'})}{\Gamma(\alpha_{\prec \tilde{m}'} + \beta_{\prec \tilde{m}'} + y_{\tilde{m}'l'})}
	\] 
	
	Computing this quantity for each situation in turn allows us to determine when and if there is a certain point where the stage is a poor forecast for the subsequent data.

	\section{Examples}\label{sec:ex}
	
	\subsection{CHDS}

	
	Other studies of the CHDS example have shown that the CEG give a much higher MAP score than the BN model/
	In this paper, we focus on the diagnostics for stratified CEG models and show how the diagnostics can be used to explain why the fit of the CEG is better.
	More explicitly, our diagnostics can be used to show where predictions from the CEG model outperform those of the BN. 
	To enable this comparison, we will compare two CEGs and the original BN.
	Figure \ref{fig:chdscegBN} shows a CEG$_\text{BN}$ that encodes additional context-specific information from previous studies \cite{CEGbook}.  
	
	The log marginal likelihood of this model is $Q(M_{\text{CEG$_\text{BN}$}})=-2,495.01$. 
	Under the relative standardization method, we obtain a Bayes Factor of 2,421,748.
	This is a tremendous improvement over the existing BN model already.
	With the assumed variable ordering $(X_{\text{s}}, X_{\text{e}}, X_{\text{l}}, X_{\text{h}})$, the AHC algorithm returns the structure CEG$_\text{AHC}$ in Figure \ref{fig:chdscegB}.
	The marginal log likelihood for CEG$_\text{AHC}$ is  -2478.49. 
	This model is an even more sizeable improvement over the original BN with a Bayes Factor of 14,946,684.
	Comparing the two CEG models, the model generated by the AHC algorithm is only six times as likely to have been data generating model, with a Bayes Factor of 6.172.
	This offers strong evidence that CEG$_\text{AHC}$ is a more suitable model for the CHDS data than the equivalent BN representation in CEG$_\text{BN}$. We will nevertheless consider both as candidate models in order to demonstrate how our monitors identify the differences in the structure.
	
	\begin{figure}[!ht]
		\centering
		\begin{subfigure}[b]{0.47\textwidth}
			\centering
			\includegraphics[width=\textwidth]{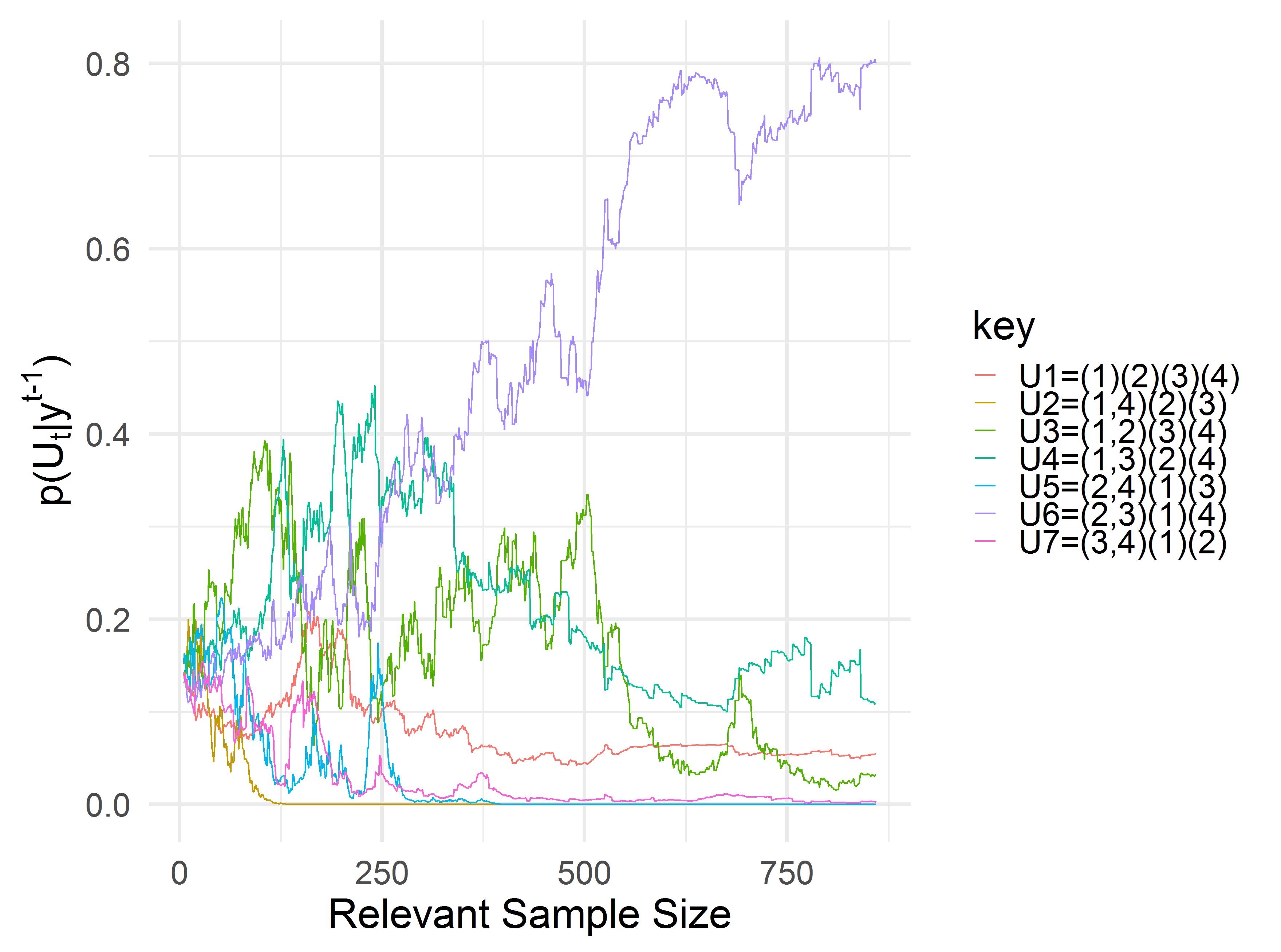}
			\caption{CEG$_\text{BN}$: the staging monitor for variable $X_{\text{l}}$. This eventually recovers the optimal staging found in CEG$_\text{BN}$.} 
			\label{fig:chdsApart}
		\end{subfigure}
		\hfill
		\begin{subfigure}[b]{0.45\textwidth}
			\centering
			\includegraphics[width=\textwidth]{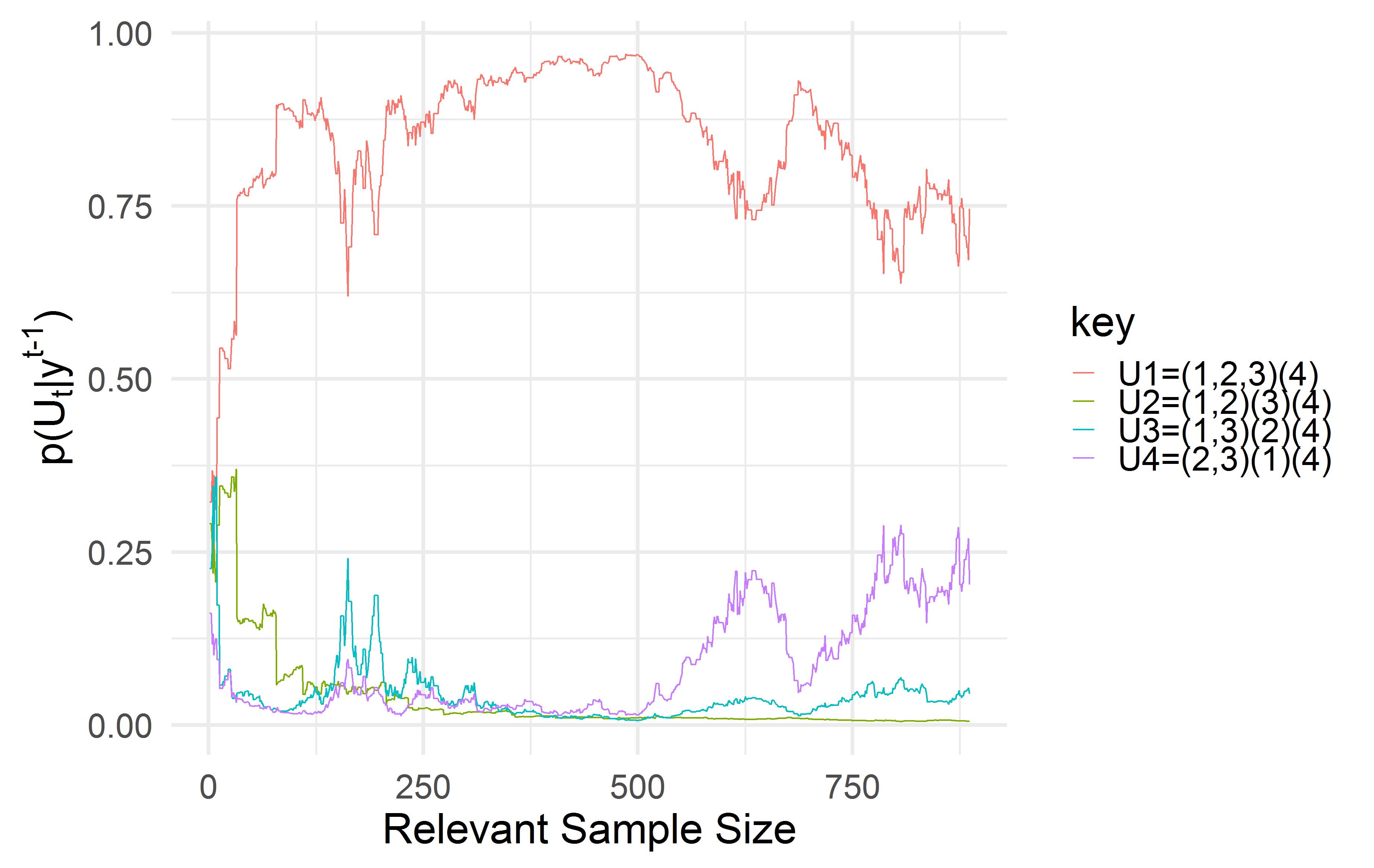}
			\caption{CEG$_\text{AHC}$: the staging monitor for variable $X_{\text{l}}$ depicting the selected staging as the most likely one.} 
			\label{fig:chdsBpart}
		\end{subfigure}
		\caption{Partition monitors for two candidate CEG models.}
	\end{figure}


	The staging monitor examines the possible partitions of the stages, called stagings at each cut in the tree.
	The staging monitor for CEG$_\text{BN}$ is shown in Figure \ref{fig:chdsApart}. It confirms that \{$X_{\text{s}} =$ High $X_{\text{e}} =$ Low, $X_{\text{s}} =$ Low $X_{\text{e}} =$ High\}, \{$X_{\text{s}} =$ High $X_{\text{e}} =$ High\}, \{$X_{\text{s}} =$ Low $X_{\text{e}} =$ Low\} (denoted (1)(23)(4) emerges as the clear preference for the staging. 
	
	We see that the model struggles to distinguish between \{$X_{\text{s}} =$ High $X_{\text{e}} =$ High, $X_{\text{s}} =$ High $X_{\text{e}} =$ Low\},
	\{$X_{\text{s}} =$ Low $X_{\text{e}} =$ High\},
	\{$X_{\text{s}} =$ Low $X_{\text{e}} =$ Low\}
	(denoted (12)(3)(4) )  and 
	\{$X_{\text{s}} =$ High $X_{\text{e}} =$ High, $X_{\text{s}} =$ Low $X_{\text{e}} =$ High\}, \{$X_{\text{s}} =$ High $X_{\text{e}} =$ High\}, \{$X_{\text{s}} =$ Low $X_{\text{e}} =$ Low\}
	(denoted (13)(2)(4)) 
	in the early observations. This suggests that an alternative model with a different stage structure might be more suitable for the data. 
	
	%

	However, the monitor for CEG$_\text{AHC}$ in Figure \ref{fig:chdsBpart}, indicates a better fit to the data. The current staging is given by \{$X_{\text{s}} =$ High $X_{\text{e}} =$ Low, $X_{\text{s}} =$ Low $X_{\text{e}} =$ High, $X_{\text{s}} =$ High $X_{\text{e}} =$ High \}, \{$X_{\text{s}} =$ Low $X_{\text{e}} =$ Low\}, (denoted (123)(4) in Figure~\ref{fig:chdsBpart}). This remains the most likely staging throughout the data. 
	

	
	\begin{figure}[!ht]
		\centering
		\begin{subfigure}[b]{0.47\textwidth}
			\centering
			\includegraphics[width=\textwidth]{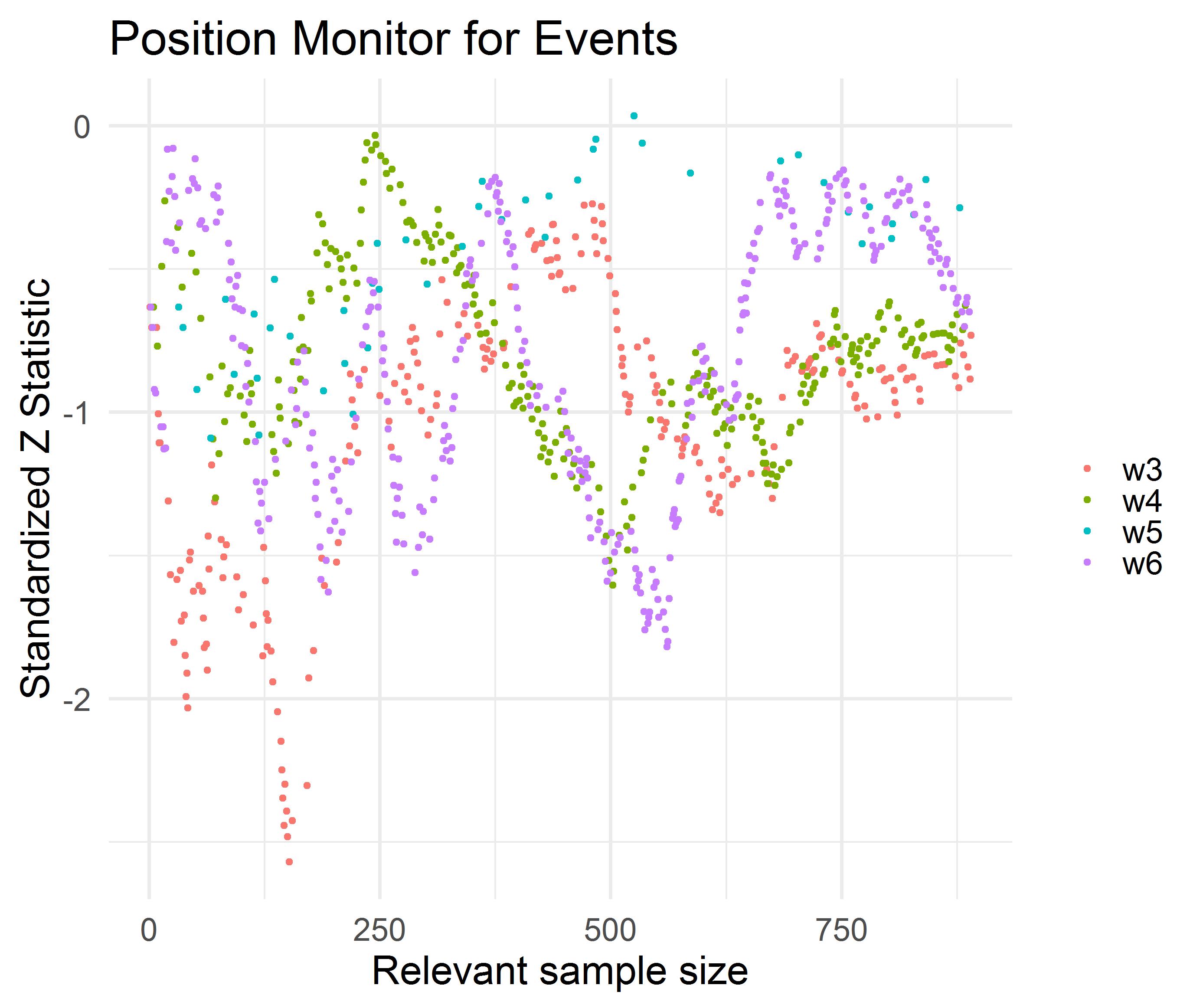}
			\caption{CEG$_\text{BN}$: the position monitor for positions $w_3$, $w_4$, $w_5$, and $w_6$ modelling $X_\text{l}$} 
			\label{fig:posCEGbn}
		\end{subfigure}
		\hfill
		\begin{subfigure}[b]{0.45\textwidth}
			\centering
			\includegraphics[width=\textwidth]{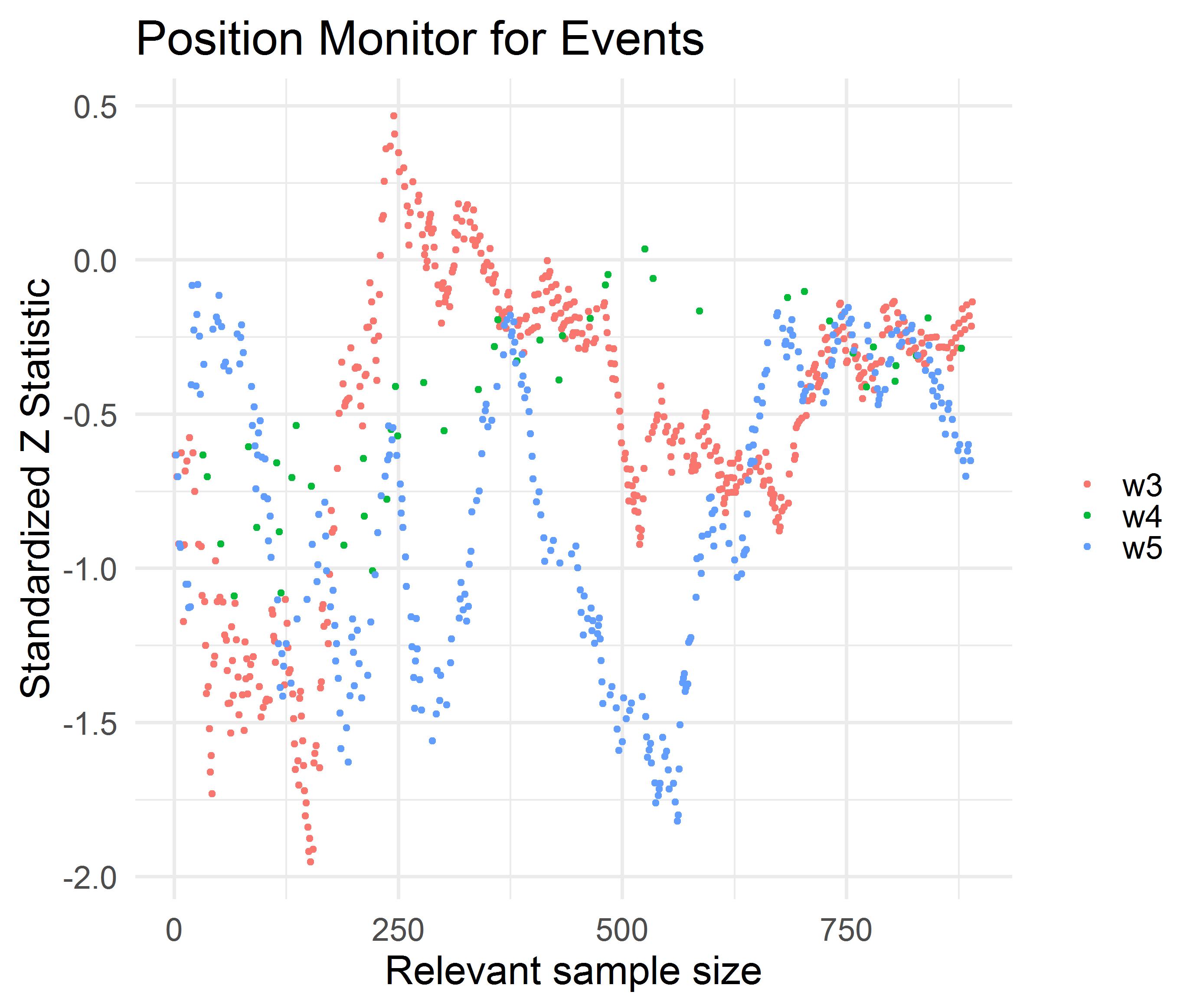}
			\caption{CEG$_\text{AHC}$: the position monitor for positions $w_3$, $w_4$, and $w_5$ modelling $X_\text{l}$} 
			\label{fig:posCEGahc}
		\end{subfigure}
		\caption{Position monitors for two candidate CEG models.}
		\label{fig:part}
	\end{figure}
	
	To confirm the more accurate modelling of positions modelling context-specific probability distributions of $X_l$ in the candidate CEGs, we can check the position monitors applied to $\text{CEG}_\text{BN}$ and $\text{CEG}_\text{AHC}$ in Figures~\ref{fig:posCEGbn} and~\ref{fig:posCEGahc} respectively. Both models are acceptable, and a substantive improvement over the position monitor of the original BN in Figure~\ref{fig:nodeBNevents}. 
	
\begin{figure*}
		\centering
		\begin{subfigure}[b]{0.25\textwidth}
			\centering
			\includegraphics[width=\textwidth]{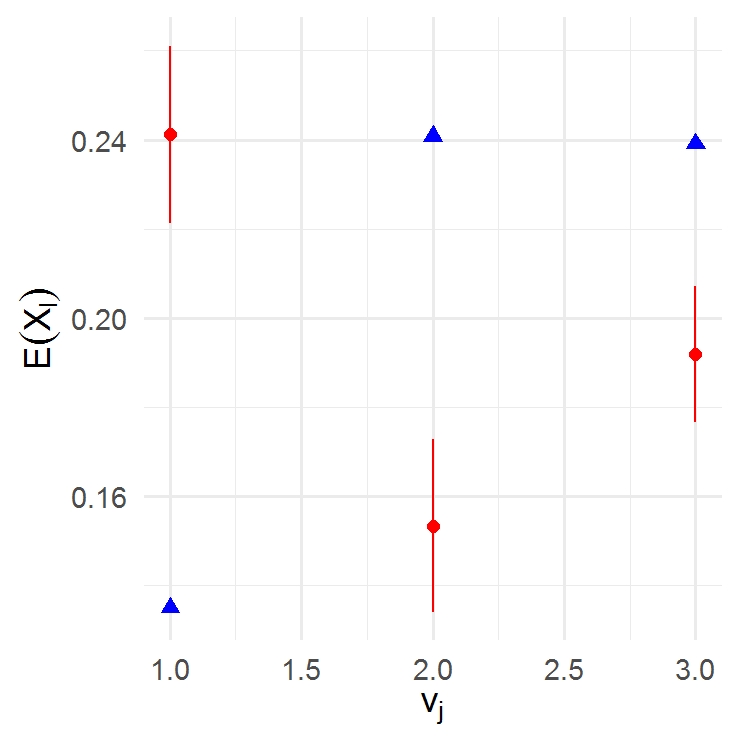}
			\caption[]%
			{{\small $u_3$ LOO monitor}}    
			\label{fig:prop3}
		\end{subfigure}
		\hfill
		\begin{subfigure}[b]{0.24\textwidth}  
			\centering 
			\includegraphics[width=\textwidth]{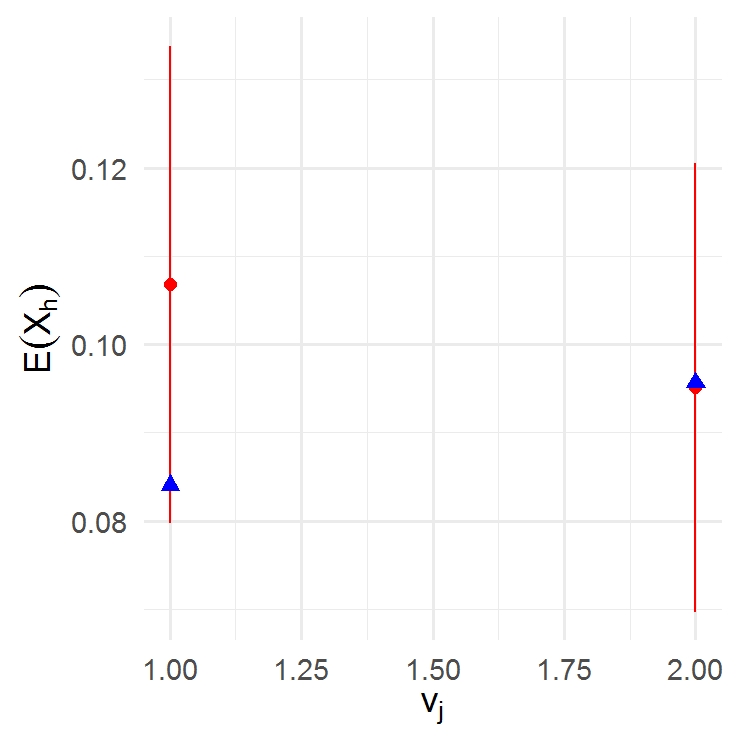}
			\caption[]%
			{{\small  $u_5$ LOO monitor}}    
			\label{fig:prop5}
		\end{subfigure}
		\begin{subfigure}[b]{0.24\textwidth}   
			\centering 
			\includegraphics[width=\textwidth]{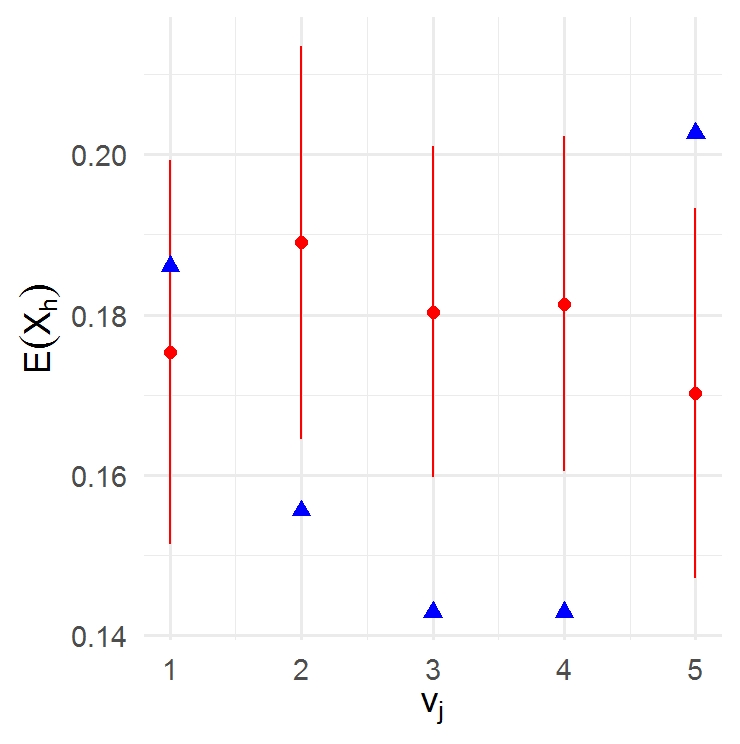}
			\caption[]%
			{{\small $u_6$ LOO monitor}}    
			\label{fig:prop6}
		\end{subfigure}
		\begin{subfigure}[b]{0.24\textwidth}   
			\centering 
			\includegraphics[width=\textwidth]{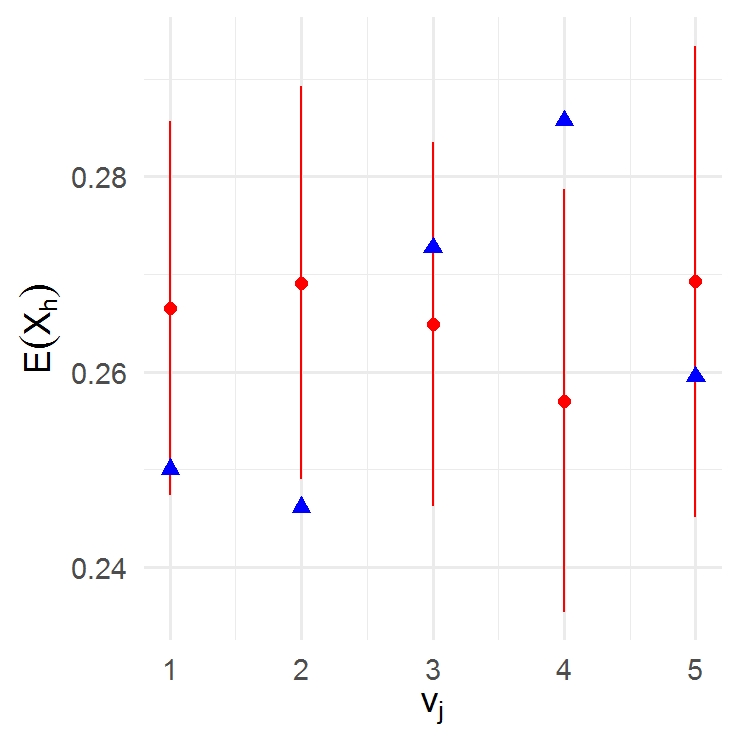}
			\caption[]%
			{{\small  $u_7$ LOO monitor}}    
			\label{fig:prop7}
		\end{subfigure}
		\caption[ ]
		{\small The observed (blue triangles) and expected (red dots) proportions of households with high adverse life events (a) and children admitted to the hospital (b,c,d) with the respective situations in Table~\ref{tab:chdsstages} removed.} 
		\label{fig:chdsprop}
	\end{figure*}

	After checking the staging, we turn our attention to the composition of the stages themselves. We consider the situations for the best-fitting CEG, $\text{CEG}_\text{AHC}$.
	In this CEG, stages $u_0$, $u_1$, $u_2$, and $u_4$ only have one contributing situation, so we examine  $u_3$, $u_5$, $u_6$, and  $u_7$. (Recall that stages are not labelled in Figure~\ref{fig:chdscegB}, but can be identified by assigning sequential labels to the unique colours.)
	The leave one out monitors return Bayes factor scores very close to zero, so we examine the plots of the expected and observed proportions of the levels of interest. We consider the proportion of $X_{\text{l}} = $ High for $u_3$ and $X_{\text{h}} =$ Yes for stages $u_5$, $u_6$, and  $u_7$ in Figure \ref{fig:chdsprop}.
	
	While the staging and position monitors for $u_3$ and $w_3$ and $w_4$ respectively suggest that the probability distribution is a good fit for the data overall, the situation monitor in Figure~\ref{fig:prop3} suggest that we should be cautious about the forecasts $\text{CEG}_{\text{AHC}}$ for families experiencing a high level of adverse events. 
	If we estimate the proportion of high adverse life events from households with either high social and low economic or low social and high economic capital, we will overestimate for households with high economic and social capital. Conversely, we underestimate the proportion of high adverse life events when we examine the leave one out proportions for $v_2$ and $v_3$. 
	
	Examining the prequential monitors here with the ordering of decreasing capital $I(X_{\text{l}}) = \{v_1, v_2, v_3\} =$  \{$X_{\text{s}} =$ High $X_{\text{e}} =$ High, $X_{\text{s}} =$ High $X_{\text{e}} =$ Low, $X_{\text{s}} =$ Low $X_{\text{e}} =$ High \} gives 
	$p(y_{2,\text{High}}) = 0.028$ and $p(y_{2,\text{High}}) = 0.102$. This further confirms that situations $v_1$ and $v_2$ are not exchangeable. 
	To adjust the model, we might consider the process by which families experience a number of life events. 
	The leave one out monitor for $u_3$ in particular suggests that something fundamentally different might be contributing to adverse life events for families with high social and high economic standing. 
	\begin{table}
		\begin{tabular}{cccc|cccc|cccc}
			$u_5$&&&&$u_6$& &&& $u_7$&&&\\
			$v_i$ & $X_{\text{s}}$  & $X_{\text{e}} $ & $X_{\text{l}} $& $v_i$ & $X_{\text{s}}$  & $X_{\text{e}} $ & $X_{\text{l}} $& $v_i$ & $X_{\text{s}}$  & $X_{\text{e}} $ & $X_{\text{l}} $  \\
			\hline
			$v_1$ & High & High & Low & $v_1$ & High & High & Average &$v_1$ & High & High  & High\\
			$v_2$ & High & Low & Low & $v_2$ & High & Low & Average &$v_2$ & High & Low & High\\
			&&&& $v_3$ & Low & High & Average &$v_3$ & Low & High & High\\
			&&&&$v_4$ & Low & High & Low &$v_4$ & Low & Low & Average\\
			&&&&$v_5$ & Low & Low & Low & $v_5$ & Low & Low & High\\
			
		\end{tabular}
		\caption{Situations composing stages modelling $X_\text{h}$ in stages $u_5$, $u_6$, and $u_7$}
		\label{tab:chdsstages}
	\end{table}

	Stage $u_5$ is composed of the situations listed in Table~\ref{tab:chdsstages}. This is the moderately fortunate group. They are characterized by low life events and high social standing. 
	The prequential monitor is given by: $p(y_{2,\text{No}}) = 0.082$, again with no evidence of a structural issue.

	
	Stage $u_6$ represents people who have access to either social or economic capital who experience an average number of life events, and families of individuals with low socio-economic standing who experience a low number of life events. This group has an average level of vulnerability. Examining the prequential stage monitors does not reveal any particular poor fits to the data: 
	$p(y_{2,\text{No}}) = 0.072$
	$p(y_{3,\text{No}}) = 0.272 $
	$p(y_{4,\text{No}}) = 0.220$
	$p(y_{5,\text{No}}) = 0.075$.

	Finally, $u_7$ represents the group with particularly unfortunate circumstances, regardless of their socio-economic stressors. All of the families of individuals reporting a high frequency of adverse life events contribute to this stage except for the group with no access to social or economic credit. 
	Again, the prequential monitors do not indicate any situations of ill-fitting structure: 
	$p(y_{2,\text{No}}) = 0.065$
	$p(y_{3,\text{No}}) = 0.243$
	$p(y_{4,\text{No}}) = 0.050$
	$p(y_{5,\text{No}}) = 0.054$.

	For stages $u_5$, $u_6$, and $u_7$, the leave one out monitors suggest that we should be cautious about forecasts for situations where the observed proportion of hospitalisations falls outside the bounds of our expected posterior. The monitors tell us which situations are over and underestimating hospitalisations. 
	
	%
	%
	%

	\subsection{Radicalisation Example}
	
	In this second example, we illustrate how our diagnostics can be applied to a much larger study. It examines the process by which individuals in a prison population are likely to be radicalised. 
	Because of the sensitive nature of this domain, the data was constructed from a simulated model based on expert judgements which were then calibrated to publicly available statistics within the UK. 
	The variables are as follows: 
	
	\begin{itemize}
		\item $X_\text{g}$ Gender: Binary variable with values Male, and Female
		\item $X_\text{r}$ Religion: Ternary variable with values Religious, Non-religious, and Non recorded
		\item $X_\text{a}$ Age: Ternary variable with values Old, Medium, Young
		\item $X_\text{o}$ Offence: Values include i) Violence against another person ii) RBT Robbery Burglary or Theft iii) Drug offence iv) Sexual offence, and v) other offence 
		\item $X_\text{n}$ Nationality: Binary variable indicating if an individual is a British citizen or a foreigner
		\item $X_\text{w}$ Network: Indicates whether the individual has intense, frequent, or sporadic engagement with known members of target criminal organisation
		\item $X_\text{e}$ Engagement: Binary variable that indicates whether or not the individual engages in radical activities.
	\end{itemize}
	
	The model was built to better explain the pathways that lead to criminal engagement. 
	So in this context, diagnostics are best used to examine how well the situations are predicting engagement in radical activities, $X_{\text{e}}$. 
	Due to the complexity and number of variables, the CEG model of the radicalisation data encodes a much richer space of causal hypotheses than the previous example. 
	A Bayes factor model selection with the AHC algorithm using the ordering assumed in the dataset returns a CEG structure with a log marginal likelihood of $-400007.3$, which we use here as a baseline to determine better fitting adjustments to the structure. 
	
	The stage partitioning for  engagement $X_{\text{e}}$ has six stages $U = \{u_{33}, u_{34}, u_{35}, u_{36}, u_{37}, u_{38}, u_{39}, u_{\text{null}}\}$ and 1080 unique positions, a much richer model. $u_{\text{null}}$ represents the stage encompassing all situations that are unpopulated.
	This is a convenient and methodologically sound way of processing the empty stages.
	A large number of situations is difficult to inspect for cohesion, so our diagnostics are particularly important here. 
	The size of each stage modelling engagement is shown in Table~\ref{tab:radicalnstages}.
	Due to the high number of situations in each stage, situations will be indexed according to their particular stage. (That is, a situation $v_1$ in $u_{31}$ is a distinct vertex from $v_1$ in $u_{35}$.)
	
	\begin{table}
		\centering
		\begin{tabular}{cc}
			stage &   $n$\\
			\hline
			$u_{33}$&     24\\
			$u_{34}$&     350\\
			$u_{35}$ &   232\\
			$u_{36}$ &    72\\
			$u_{37}$ &    112\\
			$u_{38}$ &   46\\
			$u_{39}$ &    54 \\
			$u_{\text{null}}$ & 190 \\
		\end{tabular}
		\caption{Number of situations in each of the stages modelling engagement}
		\label{tab:radicalnstages}
	\end{table}
	
\begin{figure}
		\centering
		\begin{subfigure}[b]{0.85\textwidth}
			\includegraphics[width=1\linewidth]{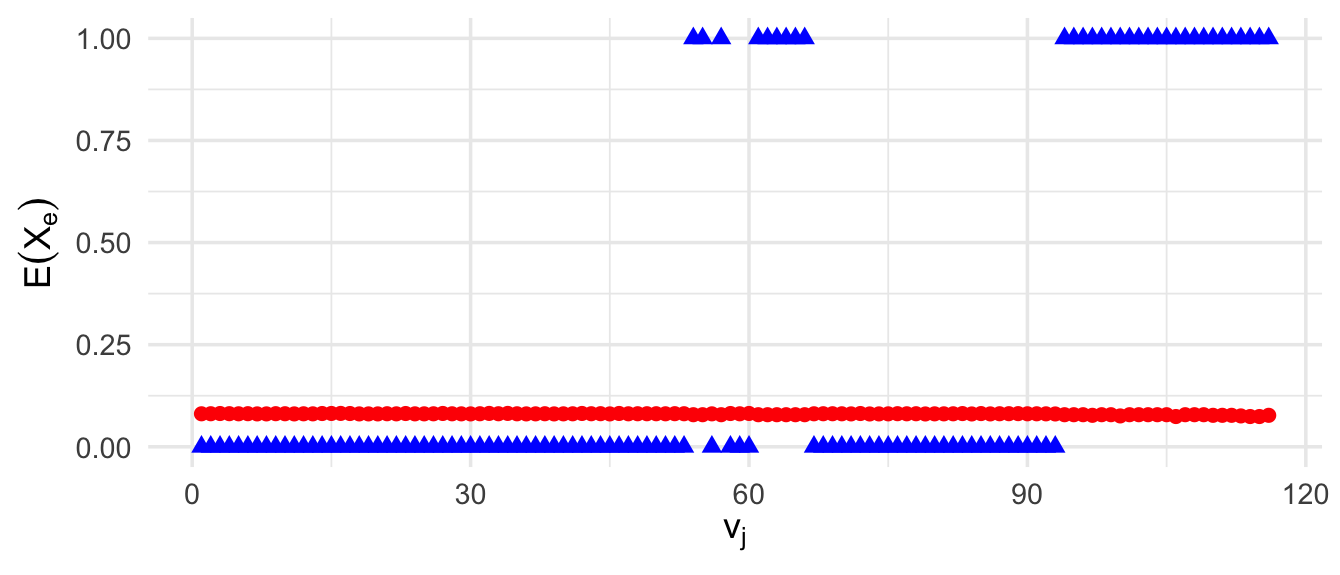}
			\caption{Leave one out monitor for $u_{35}$}
			\label{fig:loo35} 
		\end{subfigure}
		
		\begin{subfigure}[b]{0.85\textwidth}
			\includegraphics[width=1\linewidth]{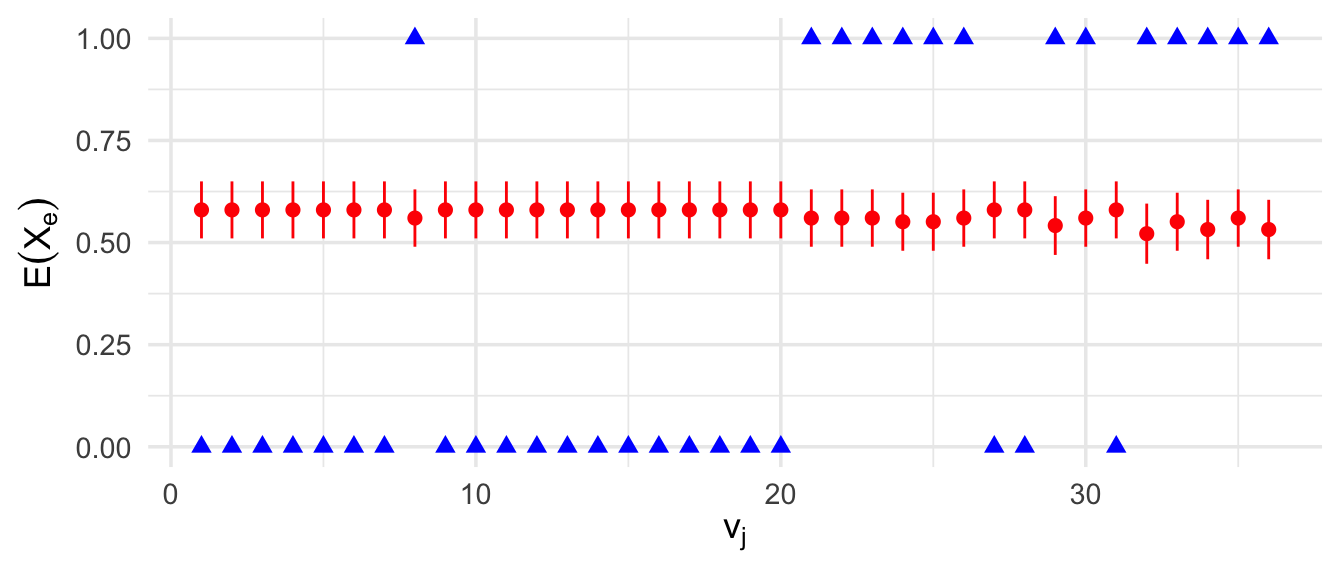}
			\caption{Leave one out monitor for $u_{36}$}
			\label{fig:loo36}
		\end{subfigure}
		
		\begin{subfigure}[b]{0.85\textwidth}
			\includegraphics[width=1\linewidth]{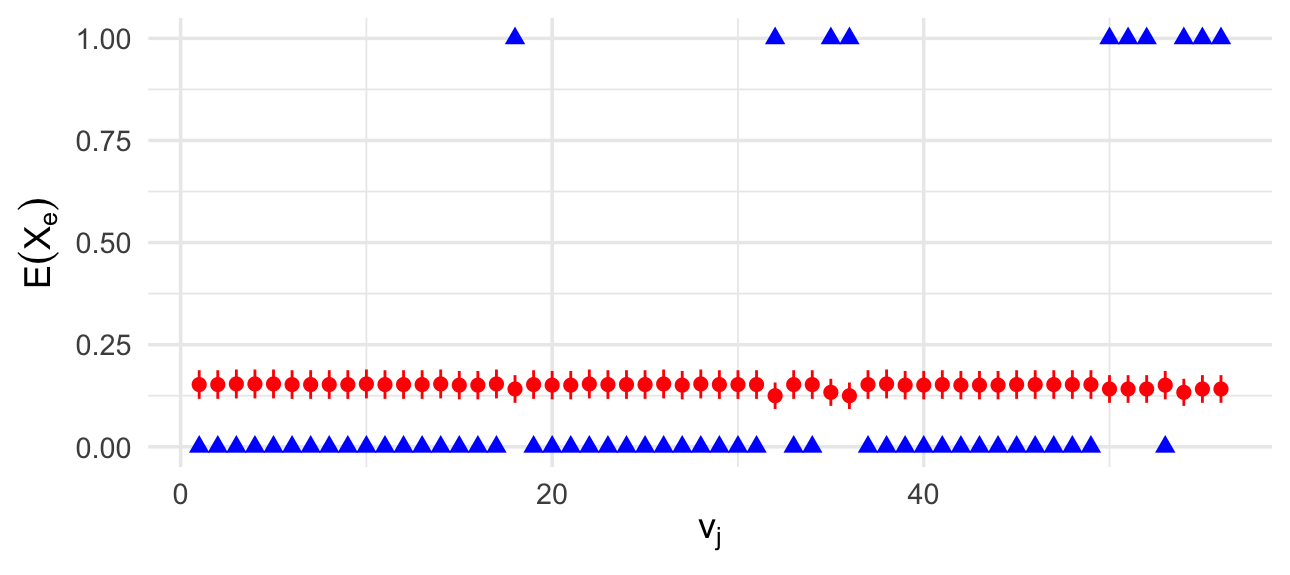}
			\caption{Leave one out monitor for $u_{37}$}
			\label{fig:loo37}
		\end{subfigure}
		\caption{Leave one out monitors for $X_\text{e}$}
		\label{fig:radicalStage}
	\end{figure}

	One of the key questions concerning the radicalisation dataset is whether or not it is accurately modelling radical engagement. 
	The stages $u_{33}$ and $u_{38}$ contains sparse situations where all engage in radical activities. 
	$u_{34}$ contains situations where no one engages in radical activities.
	$u_{35}$ contains several situations that do engage in radical activities alongside a large number of more sparsely populated situations that do not. 
	Stages $u_{36}$,  $u_{37}$  and $u_{39}$ reflect the same pattern.
	The plots of expected versus observed proportions when we leave a stage out are plotted in \ref{fig:radicalStage}.
	Plots are only shown for three stages, $u_{35}, u_{36},$ and $u_{37}$, as these are the stages that exhibit situations that exhibit both levels of $X_{\text{e}}$.

	In stage $u_{35}$ each of the situation is sparsely populated ($n < 15$ observations). Upon inspection of the dataset, the situations with all observed individuals engaging in radical activities are all religious, British males, traits not shared by the situations in which individuals do not engage. Because these counts are quite sparse and radical activity is not abundant, it is difficult to tell if the situations are exchangeable. However, the common traits seem to suggest that it would be sensible to separate out the situations representing religious British males. 
	
	In stage $u_{36}$ each of the situations is very sparsely populated ($n<5$ observations). All of the situations that have no observations that engage in radical activities have $n=1$. Thus, the expected posterior for the leave one out method has been heavily weighted by the situations containing individuals that do engage in radical activity. Again, sparsity obscures the model fit here, but inspection reveals that in $u_{35}$, the situations with observations that do engage in radical activity are all male. 
	
	This pattern holds in the last stage we consider, $u_{37}$. Again, the situations in the stage that do engage in radical activities are all male. This suggests that there is perhaps some additional information about the differences between male and female prisoners that is determining levels of radical engagement.

	This second example shows that the diagnostics are particularly useful as our model accommodates larger data sets. The changes to the situations and staging structure can be adjusted and a new global monitor computed to show that the diagnostics suggest genuine model improvements.
	\section{Discussion}
	

	Our extension of the prequential diagnostics from Bayesian Networks to the more general class of Chain Event Graphs has enables us to highlight places in which the selected structure is a poor fit to the given data.
	We have demonstrated how earlier analyses would have been much richer
	by employing these diagnostics and drawing out the reasons for certain variables failing or why one model is preferred to another.
	These monitors shown here are derived for stratified staged trees to build on the existing diagnostics available for a BN, but these methodologies also work for asymmetrical trees, a powerful example of CEG models. 
	
	These can also be applied to new classes of CEG like the dynamic CEG  \citet{Barclay2015}, where the ordering is explicit and need not be assumed from the ordering of the data.
	We have only considered models from the AHC model selection algorithm here, although we can apply these diagnostics to additional advancements in model selection criteria. 
	This work can also be extended to incorporate different score functions.
	
	The code for these CEG diagnostics as well as the subsequent two examples is available for download at \hypertarget{https://github.com/rachwhatsit/cegmonitor}{https://github.com/rachwhatsit/cegmonitor}. With the addition of the \texttt{stagedtrees} packages, we have a convenient implementation of the CEG software for practitioners. 
	
	Diagnostic monitors can be used to show how subsequent data performs when configured with the initial model. They may also be used to highlight an underlying dependence structure not captured by the existing CEG. As we have seen in the second example, the diagnostics pick out particular places for refinement as well as where situations in the model can be consolidated. 
	The prequential diagnostics shown here are a critical performance check and a useful addition to the suite of CEG methods currently available.

	\bibliographystyle{ba}
	\bibliography{CEGdiagnostics}
	
	%

\end{document}